\documentclass[]{spie}  

\usepackage{amsmath,amsfonts,amssymb}
\usepackage{graphicx}
\usepackage[colorlinks=true, allcolors=blue]{hyperref}
\usepackage{algorithm}
\usepackage{algorithmic}

\title{Smart obervation method with wide field small aperture telescopes for real time transient detection}

\author[a, b]{Peng Jia}
\author[a]{Qiang Liu}
\author[a]{Yongyang Sun}
\author[a]{Yitian Zheng}
\author[a]{Wenbo Liu}
\author[a]{Yifei Zhao}

\affil[a]{College of Physics and Optoelectronics, Taiyuan University of Technology, Taiyuan, 030024, China}
\affil[b]{Department of Physics, Durham University, South Road, Durham DH1 3LE, UK}

\authorinfo{Further author information: (Send correspondence to P.J.)\\P.J.: E-mail: robinmartin20@gmail.com}

\begin{document} 
\maketitle

\begin{abstract}
Wide field small aperture telescopes (WFSATs) are commonly used for fast sky survey. Telescope arrays composed by several WFSATs are capable to scan sky several times per night. Huge amount of data would be obtained by them and these data need to be processed immediately. In this paper, we propose ARGUS (Astronomical taRGets detection framework for Unified telescopes) for real-time transit detection. The ARGUS uses a deep learning based astronomical detection algorithm implemented in embedded devices in each WFSATs to detect astronomical targets. The position and probability of a detection being an astronomical targets will be sent to a trained ensemble learning algorithm to output information of celestial sources. After matching these sources with star catalog, ARGUS will directly output type and positions of transient candidates. We use simulated data to test the performance of ARGUS and find that ARGUS can increase the performance of WFSATs in transient detection tasks robustly.
\end{abstract}


\section{INTRODUCTION}
\label{sec:intro}  

Discovery of transients is an important scientific aim for time domain astronomy. Because transients could appear in any positions of the sky with different variation rates and different magnitudes, sky coverage, observation cadence and depth are all important for discovery of transients. However, these requirements often contradict with each other and according to different scientific observation requirements, we should make trade off between them to design different telescopes with different observation strategies. Wide field small aperture telescope (WFSAT) is a kind of telescope that can obtain images with a large field of view in a cost effective way \cite{burd2005pi, Ma2007, yuan2008chinese, cui2008antarctic, Pablo2016,ping2017the,sun2019precise}. WFSATs are widely used to detect bright transits for fast sky survey. Besides, because WFSATs are low cost, it would be possible to build a telescope array with several WFSATs to scan sky continuously, which can further increase sky coverage and observation cadence \cite{kaiser2002pan, tonry2018atlas, ratzloff2019building, lokhorst2020wide, Liu2020}.\\

During observations, huge amount of observation data are obtained by WFSATs each night. To unleash the observation ability of WFSATs and provide information for some very important targets (such as electromagnetic counterparts of gravitational wave, supernova, tidal disruption events or asteroids), these data should be processed immediately to provide necessary information for follow--up observations. Besides as large amount of transient candidates would be detected by WFSATs each night and there are only limited number of telescopes with large aperture to follow up these candidates, the accuracy of detection is important. Processing large amount of observation data to output information of transients with fast speed and high accuracy is beyond the capacity of contemporary method. New data processing method which do not need manual intervention and can output information of transient candidates with high accuracy and fast speed is required.\\

To satisfy data processing requirements of WFSATs, a lot of smart data processing methods have been proposed in recent years \cite{romano2006supernova, tachibana2018a, gonzalez2018galaxy, Burk2019astron_rcnn, Mahabal2019ML, Duev2019Deepstreaks, Jia2019Optical, Duev2019Real-bogus,2020tuprin}. These smart data processing methods use deep learning algorithms to detect or classify transient targets from observed images. In real applications, these algorithms have shown superior performance than traditional methods: they could automatically obtain positions and types of transients. For example, in the task of detecting different types of astronomical targets for fast sky survey, a faster--rcnn based astronomical target detection algorithm has better performance than that of classic methods \cite{jia2020detection}.\\ 

However, although smart data processing algorithms can improve the detection abilities of WFSATs, there are still gaps that need to be fulfilled. As we can notice that contemporary transient detection algorithms are post-processing methods, which means they process observed images afterwards. Therefore, some dim astronomical targets that would be seen by WFSATs in telescope arrays several times, but due to their low signal to noise ratio, they would not be detected by telescope arrays. Some suspicious transients could be confirmed as soon as they are detected by one of WFSATs. Detection astronomical targets from stacked images is a possible method to locate transients with low signal to noise ratio, but WFSATs in a telescope array may have different detection abilities, which makes it hard to design robust image stacking methods. Besides, stacked images would reduce detection abilities for transients with fast variations.\\

In this paper, we report a new concept transient detection framework for WFSATs in telescope arrays (Astronomical taRGets detection framework for Unified telescopes -- ARGUS). In ARGUS, each WFSAT in a telescope array will be equipped with a embedded device. An astronomical target detection neural network will be deployed in the embedded device for real--time astronomical target detection. Detection results will be sent back to data center. A classifier based on AdaBoost algorithm will be used to merge these results. Finally, the ARGUS will be able to output detection results with high accuracy for follow--up observations directly. We will report the basic structure of the ARGUS in Section \ref{sec:2} and give our preliminary results in Section \ref{sec:3}. We will make our conclusions in Section \ref{sec:4}.\\

\section{SMART TRANSIENT DETECTION FRAMEWORK FOR WFSATS}
\label{sec:2}
\begin{figure} [ht]
   \begin{center}
   \begin{tabular}{c} 
   \includegraphics[height=14cm]{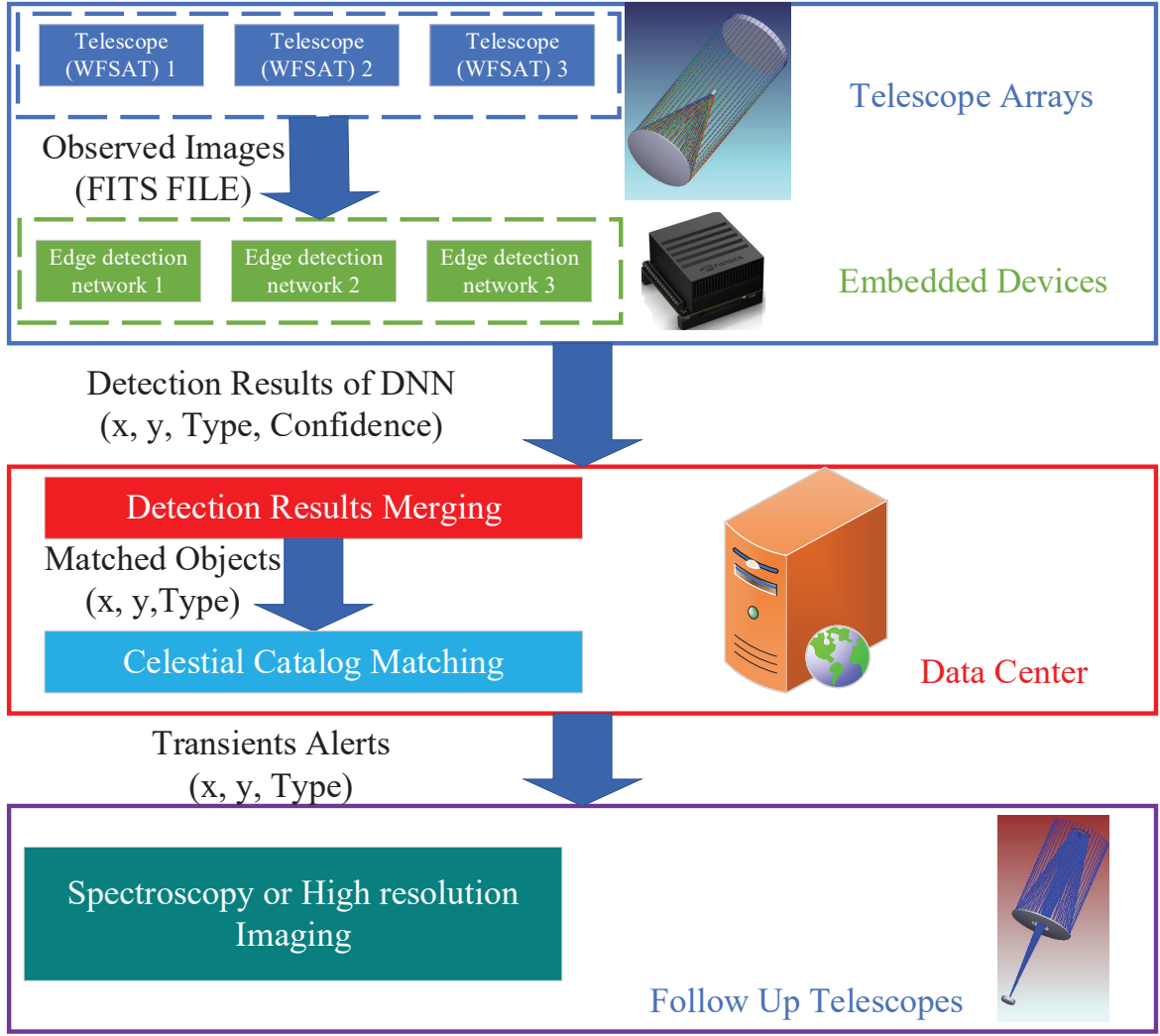}
	\end{tabular}
	\end{center}
   \caption[Framework of ARGUS] 
   { \label{fig:framework} The structure of the ARGUS for real applications. It includes two different parts that are in blue and red boxes. The remote part which includes the telescope with the embedded device is used to obtain coordinates and confidence of different astronomical targets from each WFSATs. The detection result will be sent to the data center part (red box). The data center will merge detection results and cross-match the detection results with celestial catalog. Then the ARGUS will send out transient alerts to follow-up telescopes.}
   \end{figure}

The ARGUS includes two parts: the remote transient detection part and the data center that are used to receive and process detection results. The structure of the ARGUS is shown in figure \ref{fig:framework}. We will describe each part in this Section.\\

The remote part in the ARGUS stands for the detection algorithm and the embedded device that are installed at each WFSATs. In this paper, we use NVIDIA Jetson AGX Xavier as the embedded device, because it has 512 tensor cores which can accelerate DNNs and its power cost (30 Watts) is low and its size ($105 mm \times 105 mm$) is small. A faster-rcnn based real-time astronomical target detection neural network is deployed in the embedded device \cite{jia2020detection}. Meanwhile, the detection algorithm is also being developed in a FPGA device now (F10A provided by Inspur). We have modified the structure of the faster-rcnn to make it suitable for detection of astronomical targets in WFSATs: we introduces Resnet-50 as backbone network and a feature pyramid network to extract features. The detection neural network is trained in double float accuracy in a workstation with GTX 1080 Ti Graphic Card. After training, the neural network is directly deployed in Jetson AGX Xavier also with double float accuracy. The neural network could achieve 0.5 frames per second for images with $300 \times 300$ pixels. The processing speed can further be increased if we use TensorRT to implement the detection deep neural network in the Jetson AGX Xavier and we will discuss it in our later papers.\\

\begin{figure} [ht]
   \begin{center}
   \begin{tabular}{c} 
   \includegraphics[height=8cm]{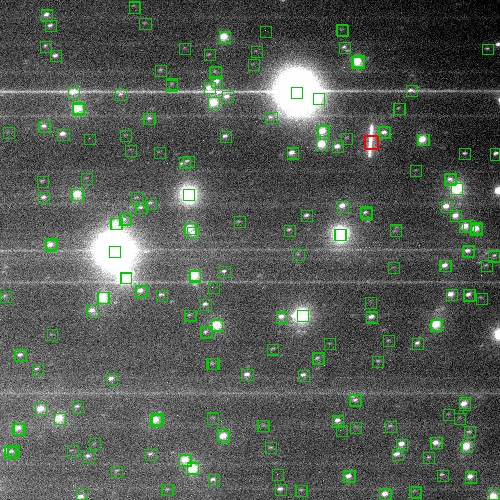}
	\end{tabular}
	\end{center}
   \caption[Detection result of the remote part of the ARGUS.] 
   { \label{fig:remotepartdet} The detection results for one frame images that are obtained by the remote part of the ARGUS in real applications. }
   \end{figure}

The faster-rcnn detection deep neural network is a stand-along astronomical target detection framework. In the last layer of a detection algorithm, there is a classification function to output positions and types of astronomical targets. One frame of detection images is shown in figure \ref{fig:remotepartdet}. Targets with green bounding box stand for stars and targets with red bounding box stand for moving targets. As shown in this figure, the detection algorithm in the remote part of the ARGUS can automatically finish source extraction and deblending tasks. Besides, the detection algorithm has better performance than classic methods as we tested with real observation data\cite{jia2020detection}. For some astronomical targets with low signal to noise ratio, the faster-rcnn would output results with low confidence. Since astronomical targets with low signal to noise ratio would be seen by WFSATs in a telescope array several times, it is straightforward to think that whether we can use detection results from several telescopes to increase detection ability of telescope arrays?\\

Therefore in the ARGUS, we try to merge detection results from each WFSATs and we have removed the classification layer and the neural network direct outputs the bounding box with confidence value in the format: $(x_1, y_1, x_2, y_2, conf)$, where $x_1, y_1$ stand for position of the bottom-left pixel, $x_2, y_2$ stand for position of the upper-right pixel and $conf$ stands for the confidence of an object being a specific target. Directly merge the detection results from several telescopes with bootstrap aggregation (average the confidence of a detection from different telescopes for a target) is a possible way \cite{bishop2006pattern}. But because image qualities will change and will introduce different point spread functions (PSF), detection results would be unstable and can affect final results. Besides, because there are a lot of WFSATs, we would be also interested in the performance of each WFSATs for astronomical targets detection tasks which could guide us to maintain instruments and optimize our algorithms.\\

To meet our requirements, we propose a dynamical classification framework in the ARGUS to merge detection results of each WFSATs. Our method is based on the adaptive boosting (AdaBoost) algorithm. The AdaBoost algorithm is used to increase the performance of a classifier through combining several weak learners \cite{schapire2003boosting}. Although individual learners is weak, the performance of the final classification model would be better, if weak learners have better performance than random guess. Based on the concept of the AdaBoost algorithm, detection algorithms from different WFSATs are assumed as a weak classifier and the ARGUS would merge these results to output detection results with high accuracy.\\

In the center part, the ARGUS includes two stages: calibration stage and implication stage. In the calibration stage, all WFSATs will observe the same sky area and their detection results will be sent to the data center. All detection results will be calibrated by the celestial catalog. Then we will modify weights of classification algorithms from each WFSATs according to the matching results between detection results and celestial catalog. The AdaBoost algorithm defined in the scikit-learn is used in this paper \cite{scikit-learn}. In the implementation stage, we will directly use weights to modify the detection results to output celestial objects. As weights of each classifier are directly related to the performance of WFSATs, WFSATs with extremely low weights will be further checked by human experts for possible malfunctions. After classification, the ARGUS will further cross-match coordinates of celestial objects with the star catalog to check for transient candidates. All celestial objects detected by the ARGUS that can not be matched by any targets in star catalog will be labeled as transient candidates. \\

\section{PRELIMINARY RESULTS OF THE ARGUS}
\label{sec:3}
In this paper, we simulate a telescope array with three WFSATs and a large sky survey telescope with the skymaker \cite{2010Bertin} to test the ARGUS. The detail information of them is shown in table \ref{tab:WFSATs}. Three telescopes (Tel2, Tel3 and Tel4) are WFSATs with diameter of 1 meter and they represent a small part of the telescope array -- Sitian \cite{Liu2020}. We set different parameters for these telescopes to test whether the ARGUS can find telescopes with best or worst performance through weights of each telescope in the data center. The first telescope (Tel1) represents the Simonyi Survey Telescope which will be used for Legacy Survey of Space and Time (LSST) \cite{tyson2002large}. Images of the same sky area obtained by these four telescopes are shown in figure \ref{fig:fourtelescopeimages}. The gray scale of these figures has been stretched by log transformation.\\ 

  \begin{figure} [ht]
   \begin{center}
   \begin{tabular}{c} 
   \includegraphics[height=10cm]{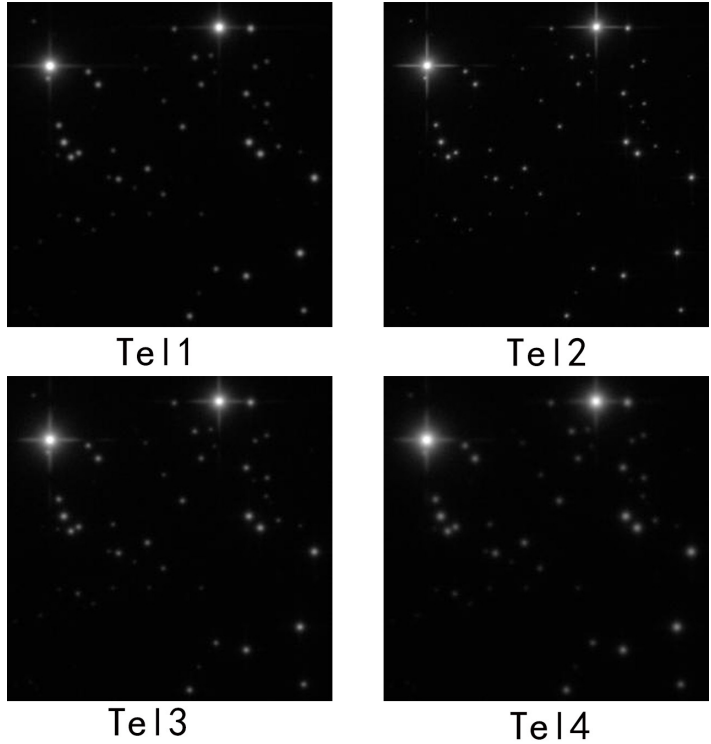}
	\end{tabular}
	\end{center}
   \caption[Images of the same sky area obtained by four telescopes. ] 
   { \label{fig:fourtelescopeimages} Images of the same sky area obtained by four telescopes. Gray scales of all images have been stretched by log transformation.}
   \end{figure} 

We use two scenarios to test the performance of the ARGUS. The first scenario is that we will use the ARGUS to merge data from both the large telescope and that from three WFSATs. The second scenario is that we will compare the performance of the ARGUS with 3 WFSATs in a telescope array and  the performance of the large telescope. Four DNN based celestial objects detection algorithms are trained in this part. We simulate observation data of each telescopes and train each DNN model with the simulated data. \\
\begin{table}[ht]
\caption{Detailed information of WFSATs used in this paper. The Seidel aberrations include defocus, spherical aberration, coma in x and y directions, astigmatism in x and y directions.} 
\label{tab:WFSATs}
\begin{center}       
\begin{tabular}{|l|l|l|l|l|} 
\hline
\rule[-1ex]{0pt}{3.5ex}  Telescopes & Tel1 & Tel2  & Tel3 & Tel4 \\
\hline 
\rule[-1ex]{0pt}{3.5ex}  Diameter (meter) & 8.4  & 1.0  &  1.0 &  1.0 \\
\hline 
\rule[-1ex]{0pt}{3.5ex}  Seeing FWHM (arcsec) & 0.7  & 0.5  & 0.9  & 1.2 \\
\hline 
\rule[-1ex]{0pt}{3.5ex} Background (mag) & 24 & 23  &  23 &  25 \\
\hline
\rule[-1ex]{0pt}{3.5ex}  Pixel Scale(arcsec) & 0.2 & 0.2  & 0.2 & 0.2 \\
\hline
\rule[-1ex]{0pt}{3.5ex} Static Seidel Aberration & 0, 0.2, 0.15   & 0.1, 0.3, 0.1 &  0.2, 0.1, 0.02&  0, 0.2, 0.1 \\
\rule[-1ex]{0pt}{3.5ex} (in waves) & 0.1, 0.2, 0.1   &  0.05, 0, 0  &  0.05, 0.1, 0&  0.05, 0.1, 0 \\
\hline
\end{tabular}
\end{center}
\end{table}

In the first scenario, we merge detection results from all 4 telescopes with the ARGUS. Because the accuracy and precision are both important for transient detection, we use f1 score ($f1 = 2*(precision*accuracy)/(precision+accuracy)$) to evaluate the performance of the detection algorithms. The detection ability will be better, if the f1 score is closer to 1. The f1 scores of the ARGUS and that of the detection results from different telescopes are shown in figure \ref{fig:fourtelescopeimages}. From these results we can find that the ARGUS could increase detection ability of telescope arrays that are composed by different telescopes. The detection results of the ARGUS are better than most telescopes (Tel1, Tel3 and Tel4). For very dim targets, its performance is slightly worse that Tel2, since it has very small FWHM (0.5 arcsec).\\

Besides, the weights of different telescopes in the ARGUS reflect their detection abilities. After training of the Adaboost algorithm in the ARGUS, the weights for each telescope are: 1.198, 1.106, 1.001, 1.018. The large telescope (Tel1) has the largest weight, because it has very large aperture to capture enough photons. The second telescope (Tel2) has the second largest weight, because it has very small FWHM to make it better in detection of dim targets. Because FWHM is defined by observation conditions of different telescopes and we can find that WFSATs with smaller FWHM have better detection abilities, it indicates us that if we place WFSATs in different regions, it would be possible to compose a telescope array that has stronger and more stable transient detection ability than any telescopes in the same region. Therefore, we test the performance of the ARGUS with three WFSATs and that of the large telescope.\\

  \begin{figure} [ht]
   \begin{center}
   \begin{tabular}{c} 
   \includegraphics[height=8cm]{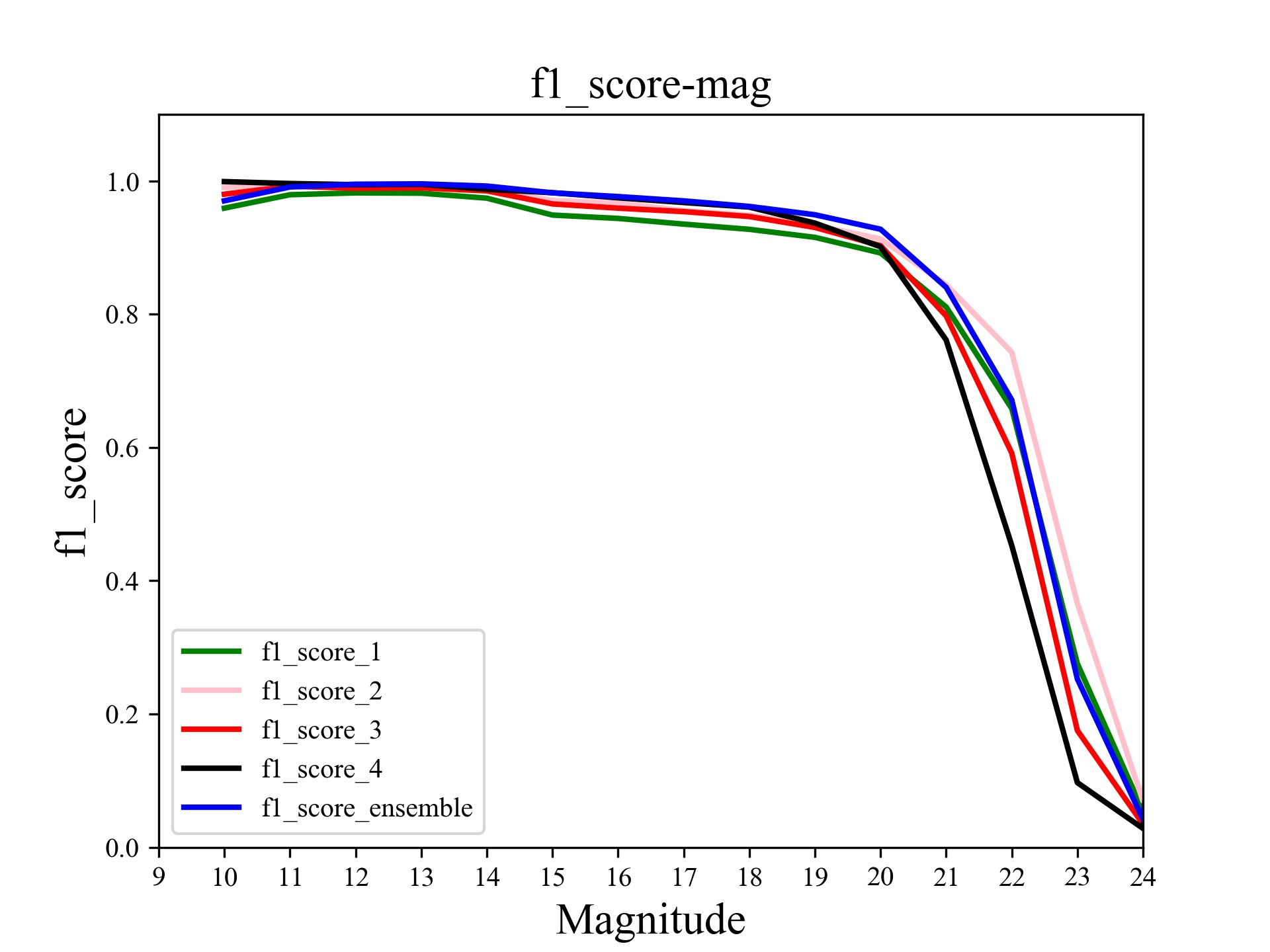}
	\end{tabular}
	\end{center}
   \caption[Detection result of the ARGUS 1.] 
   { \label{fig:fourtelescopeimages} Detection results of the ARGUS with four telescopes for celestial objects with different magnitudes. From this figure, we can find that the Tel2 has the best detection ability for dim targets ($f2\_score\_2$), while the detection ability of large telescope is affected by the larger FWHM induced by the atmospheric turbulence ($f1\_score\_2$). The ARGUS have merged all detection results and have better performance than most WFSATs ($f1\_score\_ensemble$).}
   \end{figure} 

\begin{figure} [ht]
   \begin{center}
   \begin{tabular}{c} 
   \includegraphics[height=8cm]{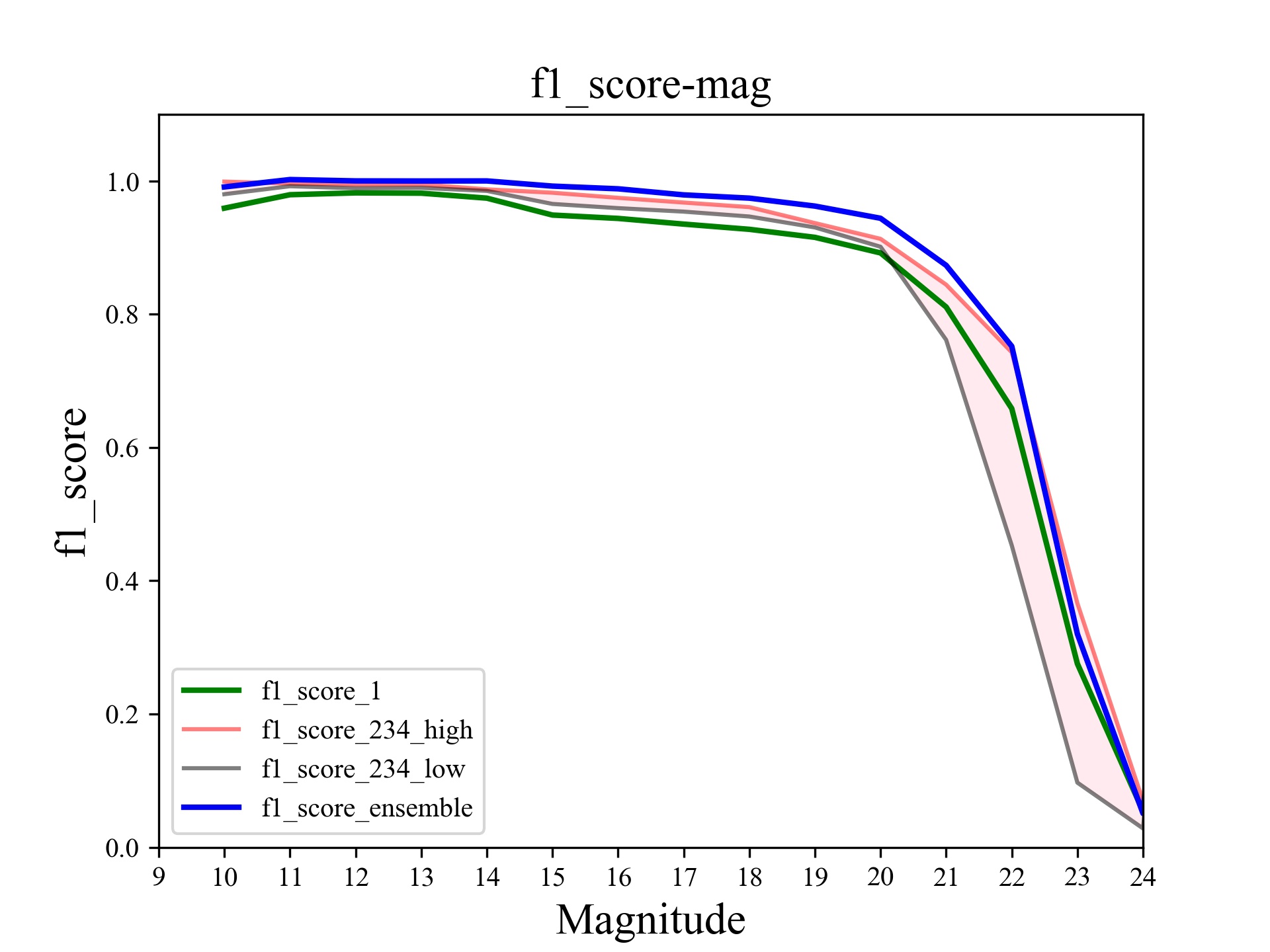}
	\end{tabular}
	\end{center}
   \caption[Detection result of the ARGUS 2.] 
   { \label{fig:threetelescopeimages} Detection results of the ARGUS obtained by three 1 meter telescopes for celestial objects with different magnitudes. From this figure, we can find that the ARGUS has better and more stable detection ability than that of any WFSATs ($f1\_score\_ensemble$ is larger than the best detection results of all WFSATs -- $f1\_score\_234\_high$). With only three telescopes, the ARGUS have better performance than that of a large telescope in detection of transients ($f1\_score\_1$).}
   \end{figure}

In the second scenario, we use the ARGUS to process images obtained from Tel2, Tel3 and Tel4 and compare detection results with that of the large telescope. The detection results are shown in figure \ref{fig:threetelescopeimages}. We can find that  the ARGUS has actively connected three WFSATs and achieves better detection ability than any WFSATs in the telescope array. It indicates us that the detection ability of the telescope array becomes better and more stable with the ARGUS. Considering WFSATs in a telescope array that distribute in different regions, the ARGUS makes the detection ability of a telescope array distributed in different regions have more chance to get better detection results than a telescope in the same region \cite{2020arXiv201102892B}.\\

\section{CONCLUSIONS AND FUTURE WORK}
\label{sec:4}
In this paper, we propose the ARGUS, a general purpose celestial objects detection framework for WFSATs in telescope arrays. The ARGUS is successor of DNN based celestial objects detection algorithms. Different celestial objects detection algorithms in different WFSATs are assumed as weak learning machine. The contribution of each WFSATs is calibrated by the calibration data and we directly use calibrated weights to merge detection results from different telescopes. The ARGUS dynamically links different telescopes for transients detection and could be used for next generation sky survey projects with WFSATs \cite{ofek2020seeinglimited}, such as the Sitian \cite{Liu2020}. In the future, we will use real observation data to optimize the ARGUS and use the ARGUS to modify observation strategy to increase observation ability of WFSATs.\\

\section{Acknowledgments}
is work is supported by National Natural Science Foundation of China (NSFC) (11503018, 61805173), the Joint Research Fund in Astronomy (U1631133) under cooperative agreement between the NSFC and Chinese Academy of Sciences (CAS). Authors acknowledge the French National Research Agency (ANR) to support this work through the ANR APPLY (grant ANR-19-CE31-0011) coordinated by B. Neichel. This work is also supported by Shanxi Province Science Foundation for Youths (201901D211081), Research and Development Program of Shanxi (201903D121161), Research Project Supported by Shanxi Scholarship Council of China (HGKY2019039), the Scientific and Technological Innovation Programs of Higher Education Institutions in Shanxi (2019L0225).  \\
\bibliography{report} 
\bibliographystyle{spiebib} 

\end{document}